# The Long Term Fate of Our Digital Belongings: Toward a Service Model for Personal Archives

*Catherine C. Marshall, Microsoft Corporation, San Francisco, California; Sara Bly, Sara Bly Consulting, Portland, Oregon; Francoise Brun-Cottan, Veri-Phi Consulting, Los Angeles, California*

## Abstract

*We conducted a preliminary field study to understand the current state of personal digital archiving in practice. Our aim is to design a service for the long-term storage, preservation, and access of digital belongings by examining how personal archiving needs intersect with existing and emerging archiving technologies, best practices, and policies. Our findings not only confirmed that experienced home computer users are creating, receiving, and finding an increasing number of digital belongings, but also that they have already lost irreplaceable digital artifacts such as photos, creative efforts, and records. Although participants reported strategies such as backup and file replication for digital safekeeping, they were seldom able to implement them consistently. Four central archiving themes emerged from the data: (1) people find it difficult to evaluate the worth of accumulated materials; (2) personal storage is highly distributed both on- and offline; (3) people are experiencing magnified curatorial problems associated with managing files in the aggregate, creating appropriate metadata, and migrating materials to maintainable formats; and (4) facilities for long-term access are not supported by the current desktop metaphor. Four environmental factors further complicate archiving in consumer settings: the pervasive influence of malware; consumer reliance on ad hoc IT providers; an accretion of minor system and registry inconsistencies; and strong consumer beliefs about the incorruptibility of digital forms, the reliability of digital technologies, and the social vulnerability of networked storage.*

## Introduction

People are beginning to accumulate significant quantities of personal digital material, material that may be meaningful over the course of their lifetimes, and in some cases, beyond [1]. Decreasing storage costs coupled with greater sophistication in consumers' abilities to create, record, obtain, and share new media has resulted in what we might think of as *digital belongings*, a mix of artifacts one has created and gathered oneself, institutional records, and published media. To a greater extent than ever before, these digital belongings form the rich backdrop of a person's life – the photos, correspondence, financial records, video recordings, the documents they read and write, their creative work, the published work they own, and much more – that come into existence in digital form and mostly stay that way.

What is the long term prognosis for this important body of digital belongings? Although we have made significant progress in our capacity to store, preserve, and access cultural heritage material, publications, institutional records, Internet material, and scientific data, our personal belongings may still be at risk; there is significant evidence that we may be heading for what Terry Kuny has referred to as a "digital dark ages" [2].

Of course, in our minds eyes, we have strategies for keeping our personal digital belongings safe: we might promise ourselves that we will track the development of new storage media, refreshing what we have already stored as needed; or we might intend to migrate our files to new formats as they become accepted standards. We might try to adhere to our own backup policies or subscribe to new services that perform IT tasks like backup for us. In fact, we scarcely notice when benign neglect takes over and we begin to rely on everyday replication tactics to keep our digital belongings safe.

In the study we report in this paper, we examine three central questions that will allow us to design a service for personal digital archiving:
- What kinds of digital belongings do people have and what do they value?
- How do people archive their digital belongings now?
- What are the central archiving challenges stemming from current practice, digital genres, and home technology environments that will guide archiving service design?

We first briefly describe our study and then go on to discuss our findings and their implications. In particular, we focus on three main themes: consumer strategies and the gaps between principles and practice; specific observed challenges for implementing a digital archiving service; and overlooked environmental factors that must also be taken into account.

## Study

We performed a field study to understand how consumers acquire, keep, and access their digital belongings with a focus on determining the extent of what they had kept, which of these belongings they cared about the most over the long term, and what obstacles they had encountered in maintaining them. Our field study consisted of three parts: an eight-interview pilot study to identify potential data collection difficulties; the main portion of the study, which included twelve in-depth interviews; and an opportunistic collection of stories about saving or recovering digital material that we gathered outside the primary interviews. All interviews were semi-structured and open-ended. The main interviews took about 1.5 hours and were conducted in the informants' homes, mostly in front of the computer(s) they used the most; other household members were sometimes present and participated in the discussions. Whenever possible, we asked our informants to show us examples of files on their computers; our tours through their digital belongings were crucial to understanding and verifying their responses to our questions.

The twelve participants (all unknown to us prior to the interviews) had each owned multiple computers and were not novice users. Ten of the twelve had multiple computers in their households. Most of them had other digital recording devices, including digital cameras, camera phones, digital video recorders, and CD or DVD burners. All had used computers for significant periods of their lives – anywhere from seven years to more than thirty. We were careful to maintain a balance of ages and genders to ensure that we could identify general factors in digital archiving. For example, several of our participants still lived at home with their families, and thus relied on a parent to maintain a portion of their personal record; others had families of their own and took responsibility for the digital belongings of others; and still others were single and childless. Participants lived in three west coast cities, which provided a further diversity of backgrounds and technological sophistication; they included several students, the owner of a pre-school, a therapist, a clerk and amateur film-maker, a partner in an IT support company, a performance artist, and a retiree who ran a small business out of her home.

The main interviews were recorded on videotape and on audiotape, augmented by still photographs. We transcribed all of the recordings for analysis and each viewed the recordings multiple times from different analytic perspectives.

## Findings

Our participants are able creators of a variety of types of digital belongings. Digital recording – through digital cameras, camera phones, and digital video – is common. In fact, some informants had blended analog and digital technologies to create interesting new forms of digital belongings: for example, one of our informants recorded video snippets from analog videotape played over her television set and recombined them, speeding them up, repeating them, and taking actions out of context; another informant had pictures from her high school classroom whiteboard taken with a camera phone. We also observed the beginnings of greater consumer participation in creating new digital document genres such as Web sites, blogs, manipulated images, complicated game characters, and IM transcripts ("convos").

Not only did participants demonstrate great sophistication about digital recording devices and applications for creating new digital genres; they also had accumulated a substantial number of digital belongings through other channels. Well-established modes of digital communication such as email have become the conduit through which digital belongings are shared: high speed Internet connections, coupled with adoption ubiquity, have turned email into a vehicle for sending and receiving photos, documents, and a variety of media clips. Most participants are comfortable with the idea of downloading material from Web sites or moving music to their computers from purchased CDs. File sharing is also common, although several of our informants became more cautious about file sharing after reading news reports about its legal implications and the added potential for virus and spyware infections.

This sophistication is not matched by a like increase in consumers' general understanding of computers and digital technology. Just as some of our informants were skilled users of complicated applications such as Photoshop and devices such as digital video cameras – and peppered their talk with a large vocabulary of specialized computer jargon ("64 bits has been hacked easy," we are told by household member seeking to protect photos of his granddaughter) – they also showed a scattered and contradictory understanding of computers in the abstract. This confusion manifested itself in what we refer to as "long cuts", patterns of computer use that demonstrate a cookbook approach to various activities. For example, the separate accounts our informant and her brother and sister had set up on a shared computer became venues for certain activities: to listen to music stored on the computer, our informant logged in as her sister. Perhaps the biggest indicator of this blend of understanding and confusion is communicated through our informants' air of fatalism and helplessness when they discuss inevitable technology changes: there is little to be done about these changes but to be swept along; just keeping up will take all we have.

The most surprising and troubling of the observed trends is that most of our informants and members of their households had already lost valuable digital belongings such as half-finished writing projects, irreplaceable photographs, and personal records. Many also had stories about computers that they could no longer use, casualties of hardware failures, security problems, or viruses. For example, as we began our interview, one informant reported, "The kids have their own [computer] upstairs which something just fried in it." Another told us, "Someone passworded the BIOS" and said he could not recover access to his files. A student explained that the family laptop had become unusable because her brother "downloads a lot of music … and all of the sudden it [the laptop] would just, like, shut down."

Certainly in business or institutional settings, such failures are not necessarily viewed as catastrophic. Systems and files can be restored through a combination of best practices and backup technology. Even so, the use of these technologies and implementation of these practices should not be confounded with real archiving; at best, they represent short term recovery solutions. However, even the most computer-savvy of our informants felt that backup would solve any long-term archiving problems, and most of the rest of the group felt that replication was sufficient. It is in the context of this common belief – that backup and file replication address long-term archiving needs – that we discuss consumer strategies for archiving and the gaps between principle and practice.

### *Consumer strategies and gaps between principle and practice*

Naturally our informants recognize the vulnerability of their digital belongings. Thus, they are inclined to develop practices to keep their digital belongings safe in the event of hardware and software calamities, accidental deletions, and malware infections. From their stories, we identified five basic strategies for archiving: (1) using system backups as archives; (2) moving files wholesale from older computers to newer computers (or to other household computers); (3) replicating specific valuable files on removable media such as CDs, DVDs, or floppy disks; (4) using email attachments as ad hoc archival storage; and (5) retaining old computers as a means of saving and accessing the files created on them. While we encountered a few instances where informants said they would print a file to save it, none thought of comprehensive hardcopy production as a viable way of keeping their digital belongings safe; hardcopy was a stop-gap when the threat was immediate or the item had already been lost. All but the last of these strategies involve file replication of some sort (either methodical or ad hoc) and many of the actual descriptions of personal strategies are, in fact, hybrids. For example, the same valuable digital photo might be sent as an email attachment, written to a CD, and included in a weekly backup, and the consumer is well aware that each copy serves as another means to recover a lost file.

Although from an archiving standpoint, none of these well-intentioned consumer strategies ensures that the digital material is actually safe, our participants conceived of them with the knowledge that benign neglect of digital materials is apt to lead to loss. These strategies are also, by and large, inconsistently implemented. The hard drive in question has not, in practice, been backed up for six months due to a recurring failure in a standard backup procedure; removable media is not available for replication when it is needed; older peripherals and removable media are no longer compatible with the current computing environment; the ad hoc IT person who acts as an intermediary between the consumer and the technology has not been over to visit for awhile; or the old hardware that has been retained specifically for its archival role no longer works.

It is instructive to examine these gaps and contradictions more carefully; they are a valuable source of insight into archiving service requirements. From the aggregated data, we identified six important folk principles that people cite regularly, principles about replicating, culling, keeping, losing, and replacing digital belongings that are often belied by everyday practice. The folk principles we describe are at times contradictory and some are patently false; however, it is important to examine them to discover the attitudes and values they reflect and the implications they have for service design.

*Replicating*. All of our informants recognized replication as a valuable safety net. In principle, it is simple and virtually free to make copies of digital belongings. One informant said, "I like to have [digital] things in two places." But, when pressed, not only was she unable to produce the removable media she planned to use as the copy's destination; she also did not know how to make such a copy herself. Another informant told us, "I could burn it on a CD but that's – I'd have to look for a blank CD somewhere." Of course, removable media is not necessary for replication, but this type of copying represented a common safety strategy. In practice, replication was often the ad hoc product of simply using the file: it was sent as an email attachment, copied to a different computer to use an application available on that platform, or published to a Web server so it could be shared.

*Culling*. Although advances in storage capacity and reductions in cost have led many technologists to advocate a "keep everything" approach to long term storage [3], consumers exhibit significant reluctance to never delete anything digital. There may be good reason for this impulse to cull: valuable material may be forgotten amid the digital dross; although storage does not impose limitations, human attention does [4]. Another factor in this reluctance stems from a fundamental confusion between active memory and storage; several of our informants equated slowness with keeping too much: "I'm, like, scared that if I do save a lot of stuff, it's going to get – it's going to slow down my computer. So. And it's just like I save the stuff that's necessary." But in other cases, culling stems from a more fundamental desire to control the technological environment; there is a tendency to delete unknown files or documents that are regarded as no longer relevant: "This could've been a seminar or something. Now I remember – what was gonna be in the seminar and I didn't go to it. Um. Wow. Yeah. I haven't looked at this stuff in a long time… [In the future] I will become a lean, mean, organizing machine."

*Keeping*. Certainly consumers have noticed the drop in storage cost and the concurrent rise in capacity. They respond by deferring decisions of whether to keep things or not and by maintaining a value-neutral stance on what they have kept. We see this dramatically played out in email, which is less frequently backed up by consumers in any methodical way than other types of files. One informant showed us a local mail file containing almost 13,000 messages that he had never backed up; another relied on server-based backup and had retained over 10,000 inbox messages, which he devalued by saying, "You want to know the truth? If I blasted my 11,230 emails away, I wouldn't be that bad off probably. Because I'd be able to work on new ones coming in." This again reflects the tension between available storage and available attention; however, when we went through some of the older messages with this informant, it was clear that some messages had been kept deliberately for their evocative power – they reminded him of people and events he would otherwise forget.

*Losing*. Accepting the possibility of incipient loss seems to be part and parcel of computer use for many consumers. They look upon their digital belongings with a wary expectation of transience. "You have to move on," one of our informants told us, comparing a recent computer crash to a house fire. Others expressed similarly existential philosophies, given the prospect of loss: "If they [my email] were totally lost it wouldn't be the end of the world. I guess I don't consider anything tangible like so important as an emotion or an experience. I guess I'm kinda like a Buddhist." Another informant, a college student, said "If my hard drive was gone, it really wouldn't bother me all that much, because it's not something I need, need. I just thought it [this paper] would be nice to keep around in case I have another assignment just like it." We noted this ambivalence about loss again, when she finds what she thinks is another missing assignment: "This is my bird CD … I never thought I'd be able to find that. It's something I just never want to throw away. I spent so much time on it." But when she discovers it is not the CD she thought it was, she tells us, "It doesn't bother me that much, only because I don't really need that [presentation] anymore. I mean, if I didn't have it, it wouldn't bother me all that much."

*Replacing*. Recent advances in retrieval have reframed desktop search as a process of re-finding, based on studies that show that frequently people are searching for something they have seen before [5]. Our informants expressed a related sentiment: the belongings that they did not create themselves were replaceable; they could be found again on the Internet, if not through a bookmark, then through a repeated search. According to one informant, a computer crash was nothing to worry about: "Nothing on here is really all that important to me, because it's all things that I could download again if I lost it." Some of our informants relied on bookmarks and favorites without realizing that these references are no guarantee against changed or missing Web sites; in fact, one person referred to her Favorites as "set in stone." However, most felt that a search would uncover the desired material – if not the same material, then something at least as good. In general, material that could be replaced (either through re-finding, re-downloading, or re-purchasing) was not regarded as important as irreplaceable material: "My pictures and my documents are more important. Because music you could always go and buy. Or you could always go and burn it somewhere else."

What do these principles and the contradictory behaviors we observed tell us? They speak volumes about value: it is difficult to state, admit, or predict the value of individual files, but consumers readily demonstrate value by what they do with a file, for example, by writing it to a CD or sending it to a friend. We also observed that sometimes it is important to be able to cull; it is central to feeling in control of one's digital belongings. It is also apparent that value is a nuanced concept that has many factors, including the personal labor and creativity that a particular digital item represents; how much emotional impact a given item has; and how hard it will be to replace, either by finding it again, reconstituting it from component parts, or by substituting something similar. We also see that sometimes it is easier to assess the value of digital assets in aggregate than it is to cull individual components; so, for example, it is easier to declare, "my email is important" than it is to assess the value of each of 10,000 messages.

Taken together, the unimplemented strategies and belied principles suggest that a service will need to be semi-automated without appearing to save too much dross or too much that is easily replaceable; that value will need to be interpreted through action and by taking a variety of important factors into account; and that an archiving service will need to be aligned with both abstract principles and with realistic practice.

### *Four Central Challenges for Personal Archiving*

Now that we know that value and benign neglect are central to the equation of what gets archived and how, we might be tempted to go off and develop a service based on these simple, yet compelling, concepts. However our study also revealed multiple types of barriers and challenges to developing such a service. We identified four challenges central to personal digital archiving: (1) digital materials accumulate in a different and more problematic way than physical materials; (2) personal digital belongings are fundamentally distributed on and among different computers, applications, and storage media; (3) standard curation problems such as managing files in aggregate, creating appropriate metadata, and migrating materials to maintainable formats are magnified in the consumer setting; and (4) facilities for long-term access are not supported by the current desktop metaphor.

*Accumulation.* As we have noted, one of the most difficult problems consumers (and archiving professionals) face is one of predicting future value [6]. Why is this of particular note in a digital environment? Why do we need to discuss the value of digital materials separately instead of bringing in best practices from physical information management? The reason is straightforward: digital belongings accumulate at a far more precipitous and unmanageable rate than physical belongings do. For example, when asked whether he got ever got rid of digital files, one participants in the pilot said: "Yes, but not in any systematic manner. And not by pruning old stuff. It's more like, I have things littering the desktop and at some point it becomes unnavigable... A bunch of them would get tossed out. A bunch of them would get put in some semblance of order on the hard drive. And some of them would go to various miscellaneous nooks and corners, never to be seen again." This challenge – the rapid accumulation of digital belongings – is formidable.

*Distribution*. A second challenge arises from how personal digital belongings are distributed; unlike many other archiving disciplines, we cannot rely on the centralization of personal assets in a single repository. Most of our informants showed us digital files that were both on- and offline, on a variety of storage media, on old and new household computers ("owned" by different members of the household), on networked email and Web servers, and sometimes on other peoples' computers that were not even network-accessible. Naturally, some of the offline files were stored on outdated media, such as Jaz or Zip drives: "I mean, they [Jaz drives] were new for, like, awhile, but then all of the sudden, you could write on CDs, so then Jaz dropped out of the picture. It was almost overnight." Often files were most accessible to their owners through their email applications and services: "I save everything [in email]. I never delete because I figure it's kind of an online journal, it's a time capsule."

*Curation*. Digital curation practices form a third archiving challenge, one that is in many ways a direct consequence of benign neglect coupled with an incomplete understanding of heterogeneous file systems and digital formats. Curation problems can be further categorized as: (1) managing files in aggregate; (2) creating appropriate metadata; and (3) migrating materials to maintainable formats. These three aspects of digital curation are well understood and thoughtfully addressed by many institutional archiving efforts. But in a consumer landscape, they are far less straightforward.

Managing files in the aggregate becomes far more difficult when the consumer does not understand heterogeneous file system structures and must handle files individually through the applications in which they were created. In other words, to them, each file is inextricably connected with the application used to open or view it. For example, when asked how she would save the elaborate graphics files she had designed on her Macintosh, one informant sighed and told us she would open them and email them to herself one by one from within Photoshop: "Well, I would go in [sighs] in here [to the file menu]. I think this is it. Then I would look – I think it's [pause] 'Save As' or 'Save for Web.' 'Attach to email.'" Unfortunately, the Macintosh in question did not have a network connection; it quickly became clear both to her and to us just how vulnerable these files were.

Benign neglect leads consumers to leave removable media unlabeled or minimally labeled; last year's taxes quickly get mixed in with a hodge-podge of commercial music CDs and application software: "I have a lot of backup here from my office when I retired… I get calls from them and they want to know something. And they can refer to it. Because I do a little bit—Ooooh! Jimi Hendrix is in there… See, this is the thing—I don't know what—so these are all of our, uh, software."

Maintaining digital assets in viable formats using strategies such as simple migration is unlikely since consumers find digital formats to be almost completely opaque. Sometimes this opacity is a direct result of well-founded efforts to hide complexity from consumers; they don't need to know about the differences in MPEG formats, only that they want to store video on a CD rather than a DVD. This opacity in turn makes consumers unsure of any other consequences of their format choices. For example, as she viewed a TurboTax document, one informant voiced her confusion, "Maybe I want to save. [reads from menu] 'Save your current tax return. Save to PFD [sic]?'" Furthermore, consumers are sometimes unaware that specialized applications such as Photoshop may need to be installed to render some files stored in application-specific formats: "some of them [photos] – it's weird – some of them came up and then I even got some of them and then all of the sudden it would die when I tried to do it." Consumers are further confused by the re-association of new display applications with existing files. One informant with a partially-installed version of FireFox told us: "Modzilla [sic] comes in for these photos for some reason. I don't know why."

*Long-term access*. In addition to archival storage and preservation, we are concerned with long-term access. Much of the work on long-term storage assumes that people will access their digital belongings through desktop search facilities or through clever visualization techniques. Automated retrieval techniques are based on a tacit underlying assumption: that is, that people basically remember what they have saved. This study – along with findings from previous studies such as [7] – suggests that human memory is fallible; our informants look for things they no longer have and have things they no longer remember. In either of these cases, many of the techniques that may be used to augment desktop search are of little help.

Nor can we extrapolate from the effectiveness of Internet search engines. Desktop search is designed for use in situations where a person is looking for something specific, something they have seen before or possibly even created themselves. The queries consumers formulate when they use Internet search engines only have to be "good enough" to answer a question or find background associated with a name, as is evident from this informant's account of helping her child with homework: "They'll say, 'okay, for Groundhog Day' – then they'll ask an obscure Groundhog Day question. Like, what does he eat? I never knew Punxsutawney Bill—Phil—ate a specific thing …" It would be dangerous to draw analogies on the effectiveness of consumer search strategies as the main means of access to stored archival digital belongings.

Although some aspects of these four challenges are documented in the archiving literature, consumer creation and use of digital materials adds many minor complications; we cannot assume consumers will get any better at assessing value as material accumulates at an ever-accelerating rate; nor can we assume any greater degree of centralization as devices and storage options and services proliferate. Formats for new media may stabilize, but the general problems associated with digital curation show no signs of abating. Finally, it is essential to investigate the requirements introduced by long-term access.

### *Important Environmental Factors*

To design a personal archiving service, it is important to understand the larger technology environment. What are the complicating factors we observed? First, we were confronted with the apparent ubiquity of viruses, spyware, and other types of malware; in fact, during one home visit, we witnessed such a failure first-hand. Second, consumers often rely on ad hoc IT support from family, friends, and other members of their extended social networks; they neither do their own IT nor call in a professional; naturally, this ad hoc support is performed with varying levels of understanding of the underlying problems. Although we tend to assume a "perfect world" when we design this sort of service, what we observed is that every one of our informants experienced an overall aggregation of minor problems on their computers, likely due to inconsistencies in the registry or partially installed software. Finally, it is important to understand underlying attitudes about digital media, technology, and the social environment in which we use them.

*Malware*. It was disturbing how many of our informants' households had fallen victim to malware; infection was common, and often misinterpreted (or stigmatized). For example, a virus acquired through file sharing was explained like this: "I think it was just too much information that we downloaded." Nor was the choice of an appropriate action to take clear: "The conundrum that I'm in is like in order to back anything up on this computer, the computer has to be working well, and in order to get the computer working well, I should have backed up everything on this computer." The over-identification of spyware by some detection software left consumers even more baffled as to the extent of the infection: "He doesn't know what she put inside the computer and how the viruses get inside, but they get inside. And he was, 'clean it, clean it, clean it." And now, now it's also some viruses… fifty-two viruses she has!" Addressing the potential presence of malware will be an important factor in service design.

*Ad hoc IT*. Although many of us have asked a more knowledgeable friend to help us with our home computers and most of us have similarly helped our friends and family in the same way, we neglect questions of agency when we design consumer services. How and when digital belongings are stored, migrated, or accessed must also take *who* into account: ad hoc IT providers are not always available and may even come into conflict with one other. Everyday maintenance may be put on hold pending the arrival of the expert. For example, one informant described a half-installed version of Firefox: "I tried to install it [Firefox] and then John [her ex-husband] said, 'Don't install anything on your computer.'… I usually defer to John. Because he's the one that's got to come over and maintain it. So I have to make sure that it's okay with him. But Jack [her 18 year old son], y'know, Jack will just do whatever he wants."

*Minor inconsistencies*. It is easy to see how the minor problems we describe may add up. But why do these minor problems matter? If we look at each of them, we see evidence of some common procedure that has gone awry; thus, in service design, we have to be wary of what we find in the registry, for example. Was the consumer *ever* able to open and view a particular file? Are items in the file system really where we expect them to be? Corroboration of important metadata values may be important as well; certainly these minor problems can lead to less trustworthy metadata.

*Attitudes*. A service design will need to take into account peoples' attitudes about their digital belongings. First of all, there is a notable optimism about the incorruptibility of digital forms in spite of experience to the contrary: "They're all digital files, why would they stop working?" Second, we encountered a considerable amount of fatalism about the reliability of digital technology; system failures are greeted by a sense that one simply needs to move on: "I think [losing digital belongings] is like losing anything else. I mean, if your house burnt down, it would hurt, kind of, and you'd just have to let go and move on." Finally, there is a fear about the vulnerability of networked digital storage to unknown social forces. This vulnerability has more to do with the violation of personal effects (e.g. photos and creative efforts) than it does about more concrete types of identity theft or financial fraud: "I don't know if I'd want to [have my] artwork, letters I read at my mother's funeral [online]… I feel more private about that than my money." These attitudes matter in designing a service that is acceptable to the people who would benefit most from it.

## Design Implications

Our findings point to both similarities to and differences from the general problem of digital preservation. Often digital preservation efforts emphasize either the long term maintenance of individual digital objects (e.g. [8]) or the creation of a centralized repository, where the objects are in canonical form (e.g. DSpace [9]). In the first case, one might look to a design that underscores a need for emulation and to maintain strong notions of provenance. In the second, one might stress ingestion process and the social aspect of providing incentives to contribute.

Guided by our four challenges (accumulation, distribution, curation, and long-term access) and our complicating environment factors (malware, ad hoc IT support, platform inconsistencies, and consumer sensitivities) we have identified four aspects of storage, preservation, and access that must be addressed by a service design.

*Layered distributed storage*. Long term storage must be designed with the idea that any centralized repository will contain both full digital objects and metadata or indices that represent digital objects held elsewhere (sometimes in long-term digital libraries and institutional stores, and sometimes in shorter-term backends such as free email accounts, personal web sites, and media-sharing venues). The architecture must also be layered to handle local storage (as it is currently distributed among local computers and devices), intermediate storage (as it is currently distributed among servers and media centers, both local and remote), and a network-based backend (which ultimately tracks distributed sources and is the final repository for unique content).

*Heuristic notions of value*. Digital assets must maintain an audit trail of value. Value may be calculated using heuristics based on at least five factors: demonstrated worth (e.g. how often an asset has been replicated), creative effort (e.g. the asset's genre and mode of creation), labor (e.g. time spent in creation), reconstituteability (in terms of an asset's source, the source's stability, and the asset's cost), and emotional impact (a factor which may be inferred by who items have been shared with). These heuristics may be used to organize stored materials for access and to guide any curation processes.

*Use-based preservation strategies*. There is a temptation to solve the most general preservation problem and commit to emulation [10]; yet evidence shows that even highly automated emulation may be costly [11]. It is more practical to store digital objects with an eye toward how they will be used later, maintaining a canonical form wherever possible [12]; some uses such as editing or custom interaction might demand emulation [13], while others will simply require that the digital asset be viewable or playable with reasonable (but possibly not complete) fidelity. For example, taking a cue from records archiving practice, PDF/A may be adequate for storing personal financial records.

*Re-encounter-based access*. Access is problematic when materials have been kept for a long time. Search is not a panacea, since you can't look for something that you don't remember you have. This study and previous studies have revealed that not only do people forget particular items they have saved; they also forget entire categories of saved material or places that they've stored treasured items. This experience of keeping valuable material in a specific well-known place – such as the box under the bed – leads us away from the desktop metaphor and into a realm of place and value.

## Conclusions

There are many who feel that benign neglect will suffice for personal digital safekeeping and that we can assume a "save everything now, decode it all later" approach to this problem. However, we have observed that consumer practices do not warrant this kind of confidence; to-date, personal losses are already significant and irreplaceable.

Each aspect of digital safekeeping – storage, preservation, and access – has its own set of entailments. Consumer assets are by no means centralized, nor do they show any sign of becoming so as people increasingly rely on a variety of networked sites (e.g. Flickr, Yahoo email) and services (e.g. online banking), where personal materials are maintained elsewhere with little guarantee of sustainability. There is an overwhelming amount of digital dross, where the replaceable and irreplaceable are irrevocably mixed, which must be sorted out before consumers feel that they are in control of the digital belongings that are personally meaningful. Many institutional or disciplinary curatorial best practices do not hold in the home environment. Finally, we cannot simply assume that our existing metaphors for saving and finding digital belongings will be serviceable over the long haul. *Our everyday digital materials, as well as our important lifetime artifacts, have proven to be significantly at risk.*

## Author Biography

*Cathy Marshall is a Senior Researcher at Microsoft; her research focuses on personal digital libraries, reading, and personal digital archiving.*